\documentclass[]{aastex63}

\received{2020 December 13}
\revised{2021 February 21}
\accepted{2021 February 25}
\shorttitle{Lorentz Reconnection Indicator}
\shortauthors{Giovanni Lapenta}
\graphicspath{{./}}
\usepackage{amsmath,amssymb}
\usepackage{mathrsfs} 
\usepackage{xcolor}
\let\tablenum\relax
\usepackage{siunitx}
\usepackage{mathptmx}

\DeclareMathOperator{\sech}{sech}
 \newcommand{\bfE}{\mathbf{E}}

\newcommand{\bfB}{\mathbf{B}}

\newcommand{\bfv}{\mathbf{v}}

\newcommand{\bfx}{\mathbf{x}}

\newcommand{\bhx}{\hat{\mathbf{x}}}
\newcommand{\bhz}{\hat{\mathbf{z}}}
\newcommand{\bhn}{\hat{\mathbf{n}}}

\DeclareMathAlphabet\mathbfcal{OMS}{cmsy}{b}{n}

\usepackage{bm}
\definecolor{granata}{HTML}{831d1c}

\begin{document}

%\preprint{}

\title{Detecting  reconnection sites  using the Lorentz Transformations for electromagnetic fields}% Force line breaks with \\

\author{Giovanni Lapenta}
% \altaffiliation[Also at ]{}%Lines break automatically or can be forced with \\
\affiliation{%
 Mathematics Department, KULeuven University, Belgium.
}
\affiliation{Space Science Institute, Boulder, Colorado, USA.}%%


%\affiliation{
%\affiliation{
%\affiliation{%


\date{\today}% It is always \today, today,

\begin{abstract}
We take a pragmatic definition of reconnection to find locations where a reconnection electric field causes a $\bfE \times \bfB$ drift that carries two components of the magnetic field towards their elimination. With this in mind as our target, we observe that such locations can be found using a new indicator: the velocity of the Lorentz transformation that eliminates two components of the  local magnetic field. Serendipitously, the indicator naturally becomes subluminal in the close proximity of a point where two components of the magnetic field vanish and it is hard zero at the vanishing location. Everywhere else the velocity of this Lorentz frame change far exceeds the speed of light. 
This property can be quickly applied in practice because computing the frame change is a local operation that requires only the knowledge of the local magnetic and electric field: it can be applied in a simulation or in observational data from a field instrument. 
We further show that the points identified can be classified in 6 categories that extend the usual types of magnetic nulls to the case of 3D reconnection in presence of a guide field. 
The approach is used to identify secondary electron scale reconnection sites in a turbulent outflow from a primary reconnection site in a highly resolved massively parallel fully kinetic particle in cell simulation. Numerous points are found and their detailed  analysis is reported. 

\end{abstract}


\section{Introduction}
Reconnection is easiest to visualize in 2D. The magnetic field in the plane reverses sign at a reconnection point and the $\bfE \times \bfB$ drift carries the frozen-in plasma toward the reconnection site where the magnetic field becomes zero right at the location of reconnection \citep{biskamp}. For its appearance, the reconnection point is called an \emph{x-point}. In this process, energy is converted from magnetic to kinetic, not just at the x-point itself but as part of the whole process of inflow and outflow.  

The x-points are easily identified in 2D as the locations where the in-plane magnetic field vanishes,  $\{B_j\}_{j=1,2}=0$, its null points. There are two types of null points, the aforementioned x-points and the o-points. The \emph{o-points} are formed in the outflow of reconnection when two connection outflow merge to form a bundle of magnetic field lines  and plasma called plasmoid \citep{finn1977coalescence} or within turbulent outflows~\citep{Fu:etal:2017}. The two types of null points are easily distinguished using the two eigenvalues of the Jacobian matrix $J_\bfB = \partial B_i/\partial x_j$ in the 2D plane ($\{x_j\}_{j=1,2}=(x,y)$) \citep{lau1990three}. We recall that the Jacobian matrix for the magnetic field must have the sum of the eigenvalues zero, since the trace of the matrix is the divergence of the magnetic field, obviously zero. We then have two possibilities: First, we have two real eigenvalues of opposite sign (with their sum zero): this is the case of the x-points. Second, the eigenvalues are complex conjugate, in this case we have an o-point. A similar classification can be obtained using the out of plane component of the vector potential, $A_z$: in this case the x and o points are obtained as null points of the gradient  $\nabla A_z=$ and the distinction between x and o points from the eigenvalues of the Hessian matrix \citep{servidio2009magnetic}.  

In 2D, the presence of an out of plane magnetic field does not complicate the story \citep{fu2016identifying} and we can distinguish \emph{antiparallel reconnection} when the asymptotic reconnecting field is entirely on the plane from \emph{component reconnection} when there is a out of plane field, called \emph{guide field}. In both cases the analysis is the same but in component reconnection the x and o points no longer are nulls in the true sense: they are only nulls for the in-plane field. 

The extension to 3D is far from easy \citep{priest-forbes}. A theoretical basis is provided by the condition required to break field line connectivity \citep{hesse1988theoretical} that can be used as a practical indicator of reconnection in 2D \citep{goldman2016can} and in 3D \citep{lapenta2015secondary} kinetic simulations. The topological equivalent to o and x points are the 3D null points, $\{B_j\}_{j=1,2,3}=0$ \citep{lau1990three} but 3D reconnection does not happen only near null points \citep{priest1995three} (a condition reminiscent of 2D anti-parallel reconnection), and more complex topological configurations are possible, for example with reconnection developing along separator lines that are lines connecting two null points \citep{parnell2010structure} (a conditions reminiscent of 2D guide field reconnection) or within flux ropes with significant twist as suggested based on  magnetospheric observations~\citep{ergun2016magnetospheric} and solar coronal modesl~\citep{torok2005confined} or typical of fusion devices~\citep{freidberg2008plasma} and expected in many astrophysical systems like jets~\citep{lapenta2006kink,li2006modeling} reproduced also in laboratory experiments~\citep{intrator2009experimental,bellan2018experiments}. 

We take here a pragmatic approach: we define reconnection, even in 3D, as a location where the $\bfE \times \bfB$ drift carries the frozen-in plasma towards a reconnection site where two magnetic field components become zero right at the location of reconnection. This definition, as it turns out, can much more easily be generalized to 3D using the Lorentz transformation of the Maxwell equations.

The Lorentz transformations of the electromagnetic field \citep{misner1973gravitation} can  always be used to find a frame where the electric and magnetic fields are parallel~\citep{landau1975classical,beklemishev1999covariant}. However, this frame change does not keep the direction of the electric field invariant. We show here that it is in fact often not possible to eliminate the components of the magnetic field in the plane perpendicular to the direction of the electric field in the laboratory frame. A real solution cannot be found and the solution becomes complex with a speed exceeding the speed of light.

The key finding of the present work is that the speed of the frame change that eliminates two components of the magnetic field drops to subluminal speed in  close proximity of a reconnection site, becoming exactly zero right at it. It is perhaps not surprising that the Lorentz frame change that removes two components of the magnetic field can be used to identify the process of reconnection defined also as the mechanism that eliminates two components of the magnetic field via an $\bfE \times \bfB$ drift. The logic is not circular because only at a site with ongoing reconnection defined as above such a Lorentz transformation exists for subluminal speeds.

The key suggestion of the present manuscript is that this property of the speed of the Lorentz transformation that eliminates two components of the magnetic field can be used as a convenient way to detect reconnection sites. Reconnection sites defined as locations where the $\bfE \times \bfB$ drift brings two components of the magnetic field to their annihilation.

We introduce then an indicator (in short called \emph{Lorentz frame indicator}) defined as the local speed of the Lorentz transformation that eliminates the two components of the magnetic field normal to the electric field and observe that this results in a very selective indicator that is lower than the speed of light only in very few locations.  The computation of this indicator is of similar complexity to the identification of null points but detects also other types of more complex 3D reconnnection topologies. The new indicator is simpler compared to other methods to find these complex 3D magnetic reconnection topologies such as the squashing factor \citep{https://doi.org/10.1029/2001JA000278, finn2014quasi} and the field integrated parallel potential \citep{lau1990three, richardson2012quasi} that requires complex integration along the field lines and even of the agyrotropy \citep{scudder2008illuminating} that requires one to compute the eigenvalues of the pressure tensor. The proposed indicator is a local operation that requires only the knowledge of the local magnetic and electric field: it can be applied in a simulation or in observational data from a field instrument. 

When a suspected reconnection point is identified,  we  show that it can be classified in 6 categories that extend the usual 2D classes of o-points and x-points. The x-points correspond to reconnection sites in their proper definition and some correspond to the o-points. Some are yet more complex 3D topologies. We show a simple  analysis of the magnetic field properties that can provide the type. 

To demonstrate the approach  proposed, we identify secondary electron scale reconnection sites in a turbulent outflow from a primary reconnection site in a highly resolved massively parallel fully kinetic particle in cell (PIC) simulation. We find numerous points and analyze the processes happening in their vicinity. We  compute the agyrotropy, the parallel electric field and study the electron flow relative to the ion flow in the vicinity.  

The manuscript is organized in Section 2 that introduces the proposed indicator of reconnection based on on Lorentz transformations. Section 3 details the classification procedure that should be used to identify the types of points found by the Lorentz frame indicator using the Jacobian of the magnetic field on the reconnection plane. Section 4 provides an overview of how the Lorentz indicator behaves in the well known and understood 2D case: this is of course a case where the new method is not really needed but it is interesting to see how it behaves where the solution is well known.  Section 5 shows the Lorentz frame indicator in the simpler but illustrative case of 2D reconnection where a local analytical formula can be obtained for the neighborhood of a point of interest.
Section 6 demonstrates the application to a massively parallel particle in cell simulation showing what reconnection points are found and analyzing their properties. Section 7  summarizes our findings and draws our conclusions.

\section{Detection based on the Lorentz transformation of electromagnetic fields}

\begin{figure}
    \centering
    \includegraphics[width=.8\columnwidth]{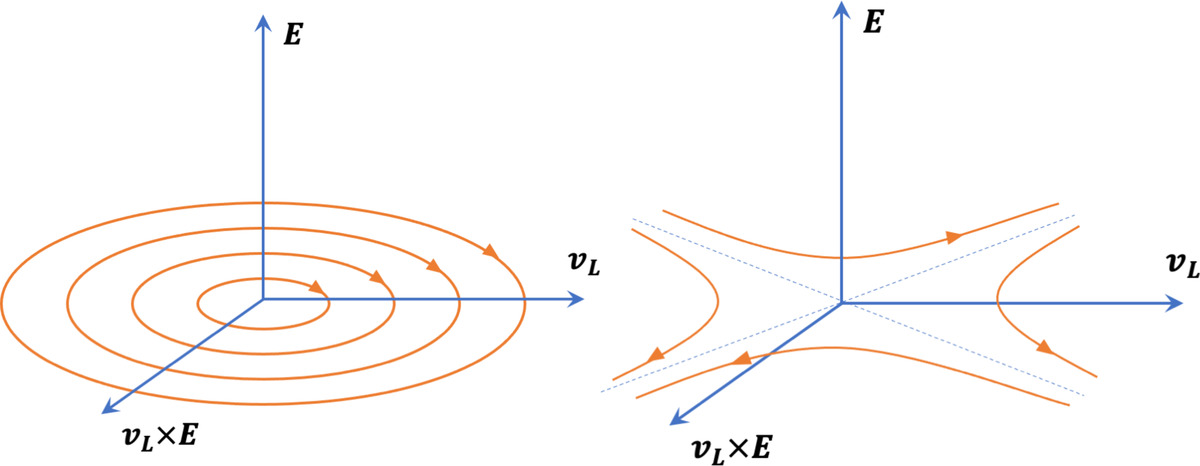}
    \caption{Orientation of the axis of the reference frame moving at speed $\bfv_L$ chosen to compute the Lorentz transformation that eliminates the magnetic field at the central location of the axis system. We choose our unit normals as: $\bhn_3 = \bfE/E$, $\bhn_2 = \bfv_L/v_L$ and $\bhn_1=\bhn_2 \times \bhn_3$. The corresponding coordinate system is $(\xi_1,\xi_2,\xi_3
)$. The example shows a o-point (panel a) and a x-point (panel b) formed by field lines laying on the $(\bhn_1,\bhn_2)$ plane.}
    \label{fig:cartoon}
\end{figure}

Let us consider one specific point, $\bfx_0$, and a specific time, $t_0$, and there define the Lorentz transformations for a frame moving with a velocity $\bfv_L = \aleph \bfE(\bfx_0,t_0) \times \bfB(\bfx_0,t_0)$, where $\aleph$ is a free scalar factor to be determined. We choose a system of coordinates in the laboratory frame that more easily highlights the direction of $\bfv_L$. We choose our unit normals as: $\bhn_3 = \bfE(\bfx_0,t_0)/E(\bfx_0,t_0)$, $\bhn_2 = \bfv_L/v_L$ and $\bhn_1=\bhn_2 \times \bhn_3$. The corresponding coordinate system is $(\xi_1,\xi_2,\xi_3
)$. Note that this choice differs for convenience from the classical choice in plasma physics of aligning the third axis with $\bfB$: in this coordinate system by construction, $\bfB$ has no component in the direction $\bhn_2$ but it has a component in the other two: $B_{\xi_3} =B_\parallel$ \emph{where parallel is meant with respect to the electric field $\bfE$} and $B_{\xi_1} =B_\bot$  \emph{where $\bot$ is meant again with respect to the electric field $\bfE$}. Figure~\ref{fig:cartoon} shows the axis choice for a case of an x-point and o-point. 

The Lorentz transformations for the electromagnetic fields in these coordinates from the laboratory frame to the frame moving with velocity  $\bfv_L$ give \citep{lorentz1904electromagnetic,misner1973gravitation}:
\begin{equation}
%\left\{ \begin{split}\mathbf {E} _{\parallel}^\prime&=\mathbf {E} _{\parallel },\\\mathbf {B} _{\parallel }^\prime&=\mathbf {B} _{\parallel },\\\mathbf {E} _{\bot }^\prime&=\gamma \left(\mathbf {E} _{\bot }+{\boldsymbol {\beta }}\times \mathbf {B} _{\bot }\right)=\gamma \left(\mathbf {E} +{\boldsymbol {\beta }}\times \mathbf {B} \right)_{\bot },\\\mathbf {B} _{\bot }^\prime&=\gamma \left(\mathbf {B} _{\bot }-{\boldsymbol {\beta }}\times \mathbf {E} _{\bot }\right)=\gamma \left(\mathbf {B} -{\boldsymbol {\beta }}\times \mathbf {E} \right)_{\bot },\end{split} \right .
\begin{split}
%B_1^\prime &= \gamma \left(B_\bot -\frac{1}{c^2}\bfv_L \times \bfE \right)\\
B_{\xi_1}^\prime &= \gamma \left(B_\bot -\frac{\aleph E^2}{c^2} B_\bot \right) \\
B_{\xi_2}^\prime &= \gamma  B_{\xi_2} =0\\
B_{\xi_3}^\prime &= \gamma B_{\xi_3} 
\end{split}
\label{Lorentz}
\end{equation}
where $c$ is the speed of light and $\gamma$ is the Lorentz factor $\gamma = 1/\sqrt{1-v_L^2/c^2}$.  The vector $\bfv_L \times \bfE$ is entirely directed along $\bhn_1$ by choice of the coordinate system (and this explains why we chose them as we did).

We can eliminate the perpendicular magnetic field component $B_{\xi_1}^\prime$ in the moving frame if we choose $\aleph=c^2/E^2$: 
\begin{equation}
    \bfv_L = c^2\frac{ \mathbf {E} \times \mathbf {B}}{E^2},
    \label{vE}
\end{equation}
as can be shown with simple algebraic manipulations substituting eq.~(\ref{vE}) into the Lorentz transformations, eq.~(\ref{Lorentz}).
In this case, by construction, the  parallel (to the electric field) magnetic field remains but the perpendicular magnetic field vanishes. This outcome is precisely what happens at a  reconnection site where the magnetic field  normal to the reconnection electric field is carried towards its elimination. 

Note that this choice of transformations differs from the textbook frame change that brings the electric and magnetic field to be parallel~\citep{landau1975classical}. In that case a frame change is always possible and it leads to the relativistic version of the $\bfE\times\bfB$ drift:
\begin{equation}
    \bfv_{rel} = \frac{\bm{\gimel} }{2\gimel^2}\left(1-\sqrt{1-4\gimel^2}\right),
    \label{vrel}
\end{equation}
with $\bm{\gimel} = \mathbf {E} \times \mathbf {B} /(E^2+B^2)$ \citep{beklemishev1999covariant}.

In our choice we restrict the transformation to eliminate the magnetic field in the plane normal to the laboratory electric field: the Lorentz transformation can become impossible because the fields become complex and the speed $v_L > c$. In fact in most practical situations, $v_L$ exceeds the typical  $\bfE \times \bfB$ drift by $B^2/E^2\approx c^2$, far exceeding the speed of light by  orders of magnitude and making this type of coordinate transformation impossible. However, this typical condition is not true in the vicinity of magnetic reconnection: there the coordinate transformation that eliminates the local $B_{\bot}$ becomes subluminal, $v_L <c$ and in fact at reconnection sites proper $v_L = 0$. 

This property of the Lorentz transformations can be understood on an intuitive basis: Reconnection is a process where a reconnection electric field causes a motion that brings the magnetic field lines and the plasma towards a region where the magnetic field energy is converted into kinetic energy and the magnetic field lines break \citep{zweibel2016perspectives}. By breaking we mean the magnetic field in the line, as it passes through a reconnection site, in that particular instant and location, disappears so that the line stops there and reforms outside the reconnection site. With this interpretation of reconnection, it is then clear that as one approaches the reconnection site the speed of the Lorentz transformation that eliminates the magnetic field (while keeping the electric field direction unchanged) locally drops and reaches zero when the field is actually annihilated in the laboratory frame.  

This is best visualized first in 2D \citep{biskamp,priest-forbes}, see Fig.~\ref{fig:cartoon}. Magnetic field lines in this case have 2 singular points, x-points and o-points. Both can be created by reconnection: the x-point is the location of magnetic field breaking and reforming and the o-points form in the outflow of reconnection when two reconnection jets form a plasmoid between two reconnection sites. The x-point and the o-point become null points of the in-plane magnetic field. They are, however, not necessarily null points as an out of plane magnetic field can be present. In this situation, Sect. \ref{sect:2dnulls} shows how the indicator $v_L$ behaves in the vicinity of the two types of points, dropping always below the speed of light in the vicinity of the singular points, being exactly zero at the x and o points. This is true regardless of the presence of an out of plane magnetic field (called guide field in the literature). In the notation used here the guide field is field component $B_{\xi_3}$ 

This peculiar property of the Lorentz transformation becomes especially convenient as a reliable indicator of reconnection in the generalisation to 3D dimensions where component reconnection in absence of true null points cannot be easily identified.  %As note above in 2D there is a standard well tuned way to recognise x and o points. %In 2D, there are standard topological ways to define the x and o point: the component out of plane of the vector potential, $A_z$ provides a direct measure of their location. The singular points are neutral points of $A_z$ where $\nabla A_z=0$. The eigenvalues of the local Hessian matrix determine the type of point\citep{servidio2009magnetic}:  If the eigenvalues have the same sign, an o-point is present. If the eigenvalues have opposite signs, an x-point is present. %Note that an eigenvalue cannot be zero because the magnetic field has zero divergence. 

%In 3D, the situation is much more complex. The direct extension ox x and o points are nulls \citep{lau1990three}. Nulls are, however, not the only topological places for reconnection to happen. A more natural extension of x-points in 3D are separator lines connecting two nulls \citep{parnell2010structure}. 

%We focus here on one aspect of reconnection more suitable for kinetic regimes of reconnection. 

The Lorentz transformation-based indicator makes the task easy. %In kinetic reconnection,  a reconnection point does not stay a point but rather tends to expand along the direction of the reconnection electric field \citep{huba2002three,shay2003inherently,lapenta2006kinetic, shepherd2012guide, li2020three}. In fact the motion of the electrons and ions around the reconnection region tends to form an elongated structure but the overall evolution still presents a reconnection electric field that carries the magnetic field towards the reconnection site where it is annihilated to reform new connectivity for the field lines.   
Finding these regions becomes  simply the task of identifying where the  $v_L$ indicator drops to subluminal speeds, and to zero at the reconnection site proper. 
When we consider the $v_L$ indicator in 3D, the velocity of the Lorentz transformation that brings the magnetic field to zero is again dropping to subluminal speed in the vicinity of true 3D nulls and  hard zero  right at null points proper (for the obvious reason that the laboratory frame already has zero magnetic field). 
But the speed $v_L$ drops to subluminal and zero also at separators or at any location of our pragmatic definition of reconnection, i.e. at locations where the $\mathbf {E} \times \mathbf {B}$ brings two  magnetic field components to be annihilated, regardless of the size of the  field in the third direction (guide field). 

Short of studying the whole topological connectivity of the magnetic field~\citep{bungey1996basic,priest19973d}, there is no simple way to determine the presence of reconnection happening without nulls. Nulls can be detected in 3D but other types of reconnection require to track the full complexity of the magnetic topology in 3D. This is obviously not possible in local measurements taken from a spacecraft and it is highly inconvenient even in simulations.
The Lorentz frame indicator $v_L$, instead, is a local measure that can be computed point by point and drops to subluminal speeds in the vicinity of any 3D feature extending to 3D component reconnection. The Lorentz frame  indicator can then easily identify the 3D equivalent of   an  x-point (the x-line) or  an o-point (a flux rope).  At the actual topological singularity of the magnetic field, the indicator becomes hard zero. The latter is, of course, unlikely to happen exactly at any point on a computational grid even in a highly resolved simulation. But locating the exact point where $v_L$ is hard zero is not needed in practice because a whole finite size region around an x or o point (and their 3D extension) become subluminal, greatly facilitating the detection in simulations and observations.

\section{Classification of 3D Lorentz subluminal regions}
\label{sect:classification}
Once a region of subluminal $v_L$ Lorentz frame indicator is found, the determination still needs to be made as to whether it is a x-point or o-point like feature. This task is easy in 2D. As noted above, in this case the Jacobian has zero trace, i.e. the magnetic field has zero divergence. X-points correspond to a Jacobian matrix with  two real eigenvalues (they forcibly have  opposite sign because the trace of the matrix, corresponding to the sum of the eigenvalues, is zero). O-points correspond to a Jacobian matrix with  complex conjugate eigenvalues. 

The situation in 3D is more complex. The Lorentz frame indicator finds a point where a subluminal transformation eliminates two components of the magnetic field and a direction $\bhn_3$ for the reconnection electric field. In the perpendicular plane $(\bhn_1,\bhn_2)$, the electric field causes the $\bfE\times\bfB$ drift that conveys the magnetic field lines projected in the pane $(\bhn_1,\bhn_2)$   to their break and reconnection. However there is an out of plane magnetic field, $B_{\xi_3}$, called guide field and aligned with the reconnection electric field. The in-plane $(\xi_1,\xi_2)$ field than does not have zero divergence: the full 3D field has zero divergence, including the contribution of $\partial B_{\xi_3}/\partial \xi_3$.

\begin{figure}
    \centering
    \includegraphics[width=\columnwidth]{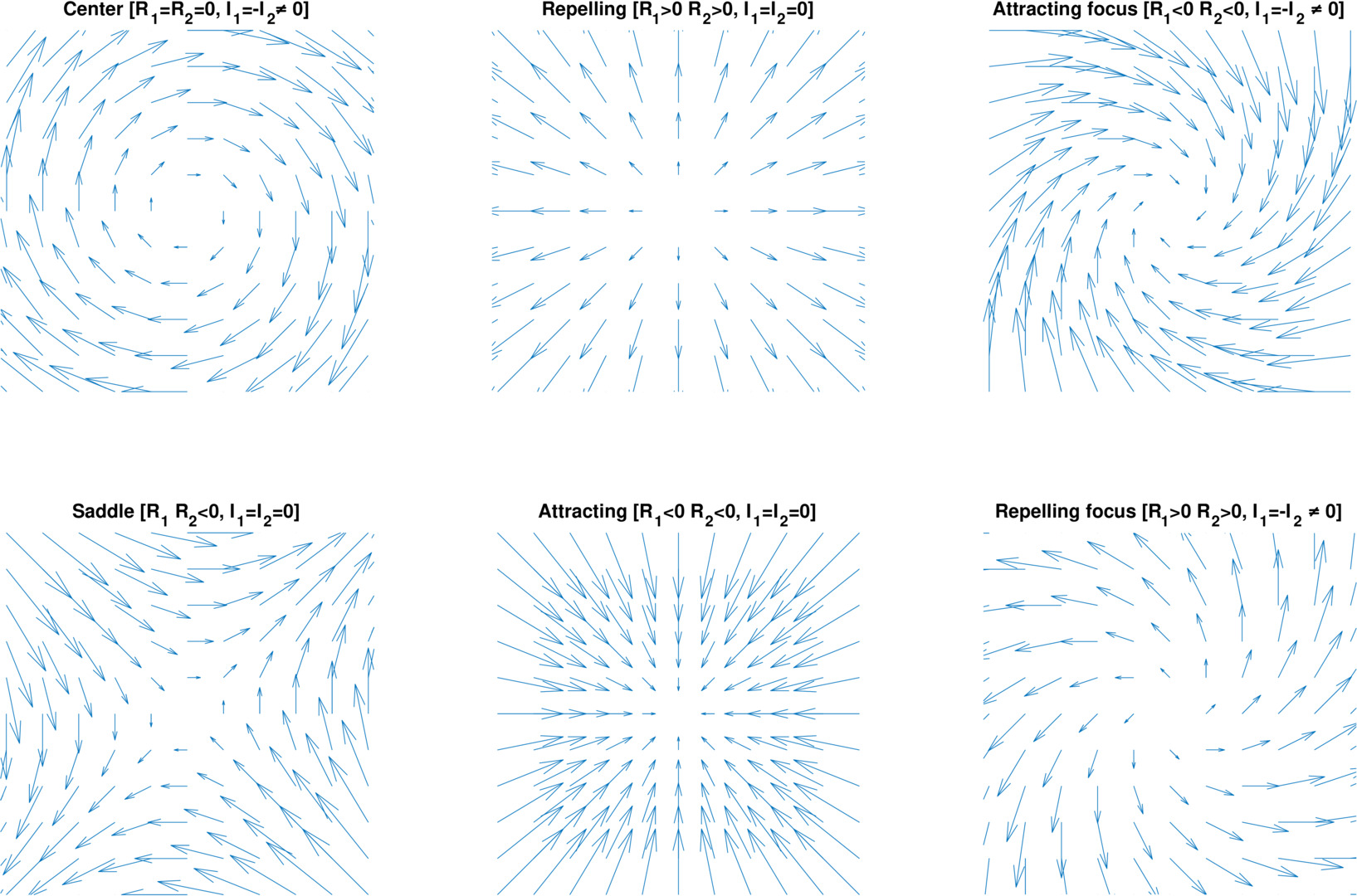}
    \caption{Six possible configurations around a null of  a generic vector field on a 2D plane. The six configurations depend on the eigenvalues of the Jacobian matrix.  For each type of 2D null, the real and imaginary part of the two eigenvalues of the Jacobian matrix are indicated and the standard name is given as commonly used in the relevant literature \citep{helman1989representation}. }
    \label{fig:vect_field}
\end{figure}

The vector field in the $(\bhn_1,\bhn_2)$ plane has then a more complex structure. Around a in plane null a generic 2D vector field (not endowed with a zero divergence) has  6 different possibilities \citep{helman1989representation} shown in Fig.~\ref{fig:vect_field}.

From the eigenvalues of the Jacobian matrix on this 2D plane $J_{B_{\xi_1,\xi_2}} = \{\partial B_{\xi_i}/\partial \xi_j\}_{(i,j=1,2)}$, we can find for each location flagged by the Lorentz frame indicator what is the type of behavior of the vector field. 

The possibility for non-zero divergence of the vector field on the 2D plane $(\bhn_1,\bhn_2)$ has increased the diversity of behavior. We recognize immediately the closest relatives to the x-point as the saddle and of the o-point as the center. These are the only two points still possible in a divergence free field. The other 4 classes of points are  allowed by the non zero 2D divergence (compensated for by the derivatives in third dimension).  These points have a configuration that is reminiscent of that of spine and spiral nulls but they are not  nulls in 3D in the sense discussed in \citet{greene1988geometrical,lau1990three}. The types of reconnection uncovered by the Lorentz transformation include all the nulls but extend the set to many more cases with more complex 3D topology. 

\section{Speed of the Lorentz frame eliminating the magnetic field in the vicinity of a 2D magnetic topological singularity}
\label{sect:2dnulls}
Before considering the 3D full complexity, it is educational to consider first the well known 2D case where we know by immediate visual inspection where an x point or an o point is, just by looking at the out of plane component of the vector potential (aka the flux function). It is useful to explore in 2D the vicinity of a x-point and o-point  the  behavior of the velocity required by the Lorentz transformations to eliminate the local value of the magnetic field.

When we require the 3D electric and magnetic fields to depend only on the 2D variables $x,y$ (sometimes this condition is referred to as 2.5D), the topological conditions of reconnection become rigorous and can be easily characterized using the out of plane vector potential component $A_z(x,y)$. In the vicinity of an o-point of x-point, a similarity solution can be constructed to represent the local magnetic  topology \citep{imshennik1967two}:
\begin{equation}
    A_z(x,y) = \frac{B_0d_i}{2}\left(\frac{x^2}{d_x^2} \pm  \frac{y^2}{d_y^2} \right)
\end{equation}
where we assumed the location of the magnetic topological feature in the origin of the coordinates. The sign between the two terms determines whether the magnetic point is of type o (+) or x (-). The values of $d_x$ and $d_y$ determine the axis of the hyperbolic (x-point) or elliptic (o-point) magnetic field lines. In the case of o-points, their ratio determines the aspect ratio of the flux rope surrounding the point. In the case of x-point, the ratio determines the reconnection geometry and is typically of order 10, a consequence of the well known "universal" fast kinetic reconnection rate \citep{shay1999scaling,birn-priest}.

The magnetic field can then be computed simply as $\bfB = - \bhz \times \nabla A_z  + \bhz B_g$. In presence of a local reconnection electric field $\bfE =  \bhz E_r $ (assumed uniform as required for steady reconnection) one can compute easily the Lorentz frame indicator as:
\begin{equation}
    v_L(x,y) = c B_0d_i\frac{\sqrt{x^2/d_x^4 +  y^2/d_y^4}} {E_r}
    \label{eq:appvl}
\end{equation}

\begin{figure}
    \centering
    a)\\
    \includegraphics[width=\columnwidth]{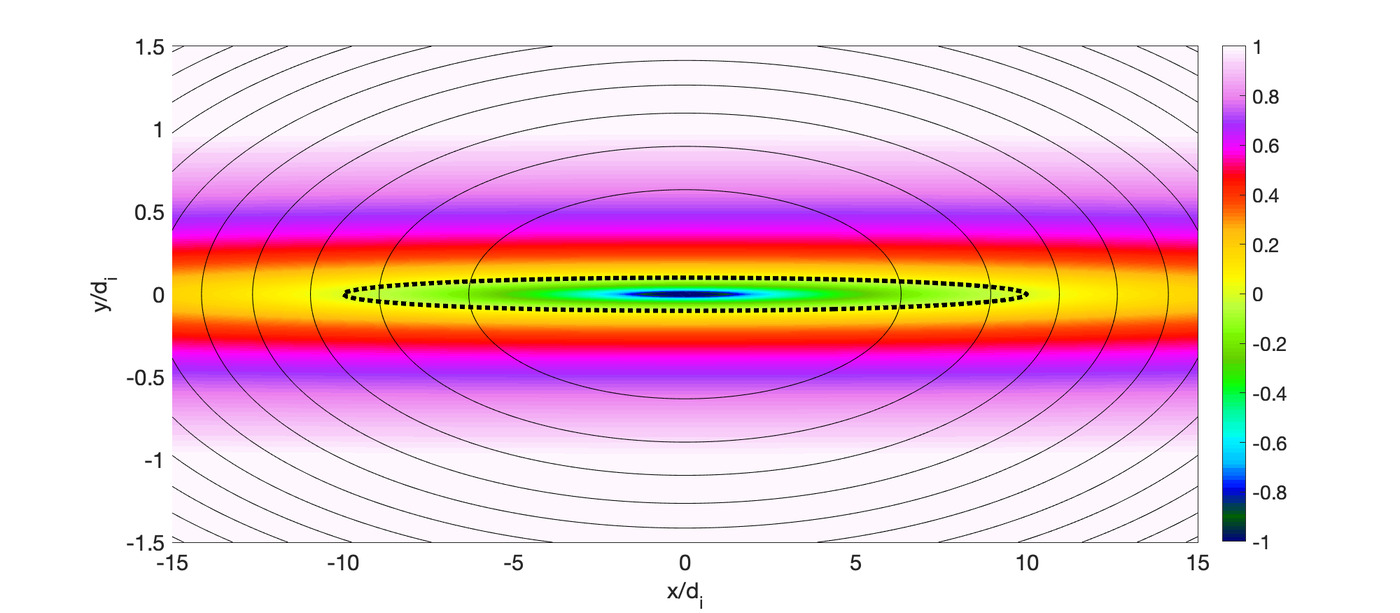}\\ b)\\
    \includegraphics[width=\columnwidth]{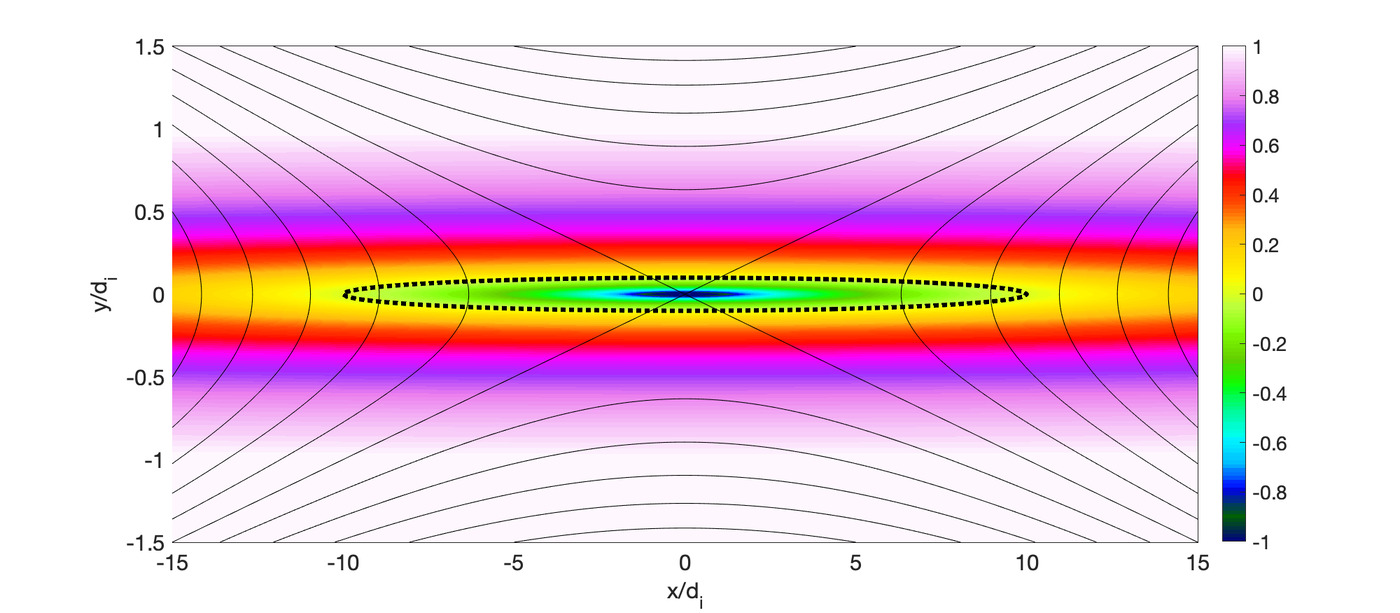}
    \caption{Speed of the Lorentz transformation that eliminates the local field. The speed is normalized to the speed of light $c$. The dotted line correspond to the frame speed equal to the speed of light. The solid black lines are the magnetic field lines. The case of the o-point (panel a) and x-point (panel b) are shown. The Lorentz frame indicator is identical in both cases.}
    \label{fig:analitiche}
\end{figure}

Figure \ref{fig:analitiche} shows the results for a typical choice of $d_x=10d_i$, $d_y=d_i$ and $E_r/v_AB_0=0.1$. As can be seen, in the central o or x point the speed of the Lorentz transformation is zero but it increases away from it. The speed becomes luminal at an ellipse around the point: all points within the ellipse $v_L=c$ have subluminal Lorentz transformation speed and can be used to detect the vicinity to the o or x point.  This is an important property because it is hard to find the hard zero for $v_L$ but it is much simpler to determine areas where the speed becomes subluminal. Figure \ref{fig:analitiche} shows these regions to be tightly confined around x and o points.
From the perspective of the Lorentz frame indicator there is no difference between o and x points.

Note that this derivation assumed the presence of only a out of plane  reconnection electric field. This is true only in a small region of electron skin depth size in the vicinity of the reconnection point. At larger distances, there is an in plane electric field \citep{chen2011inversion}.% and the Lorentz frame indicator becomes superluminal. %that modifies the conclusions above and moves the frame speed out of the plane invalidating the assumption of the Lorentz frame indicator. 
%For this reason, the Lorentz frame indicator is suitable for detecting electron scale topological features of the magnetic field. 
The extension to any orientation of the electric field is straight-froward. In 2D, the process of reconnection is constrained by the geometry to remain within the 2D plane. We need now to define the Lorentz transformation that eliminates the in-plane field only: $\bfB_{2D} = - \bhz \times \nabla A_z $:
\begin{equation}
    \bfv_L = c^2 \frac{\bfE \times \bfB_{2D}}{E^2},
\end{equation}
and the modulus of the frame speed is computed as:
\begin{equation}
    v_L = c^2 \frac{\sqrt{E^2B_{2D}^2- (\bfE \cdot  \bfB_{2D})^2}}{E^2}
\end{equation}
that is obviously zero when a point is a null for the in plane magnetic field. 
The 2D case for the Lorentz frame indicator is therefore directly related to the standard method based on finding the nulls of the in plane field~\citep{servidio2009magnetic}. %The useful application of the Lorentz frame indicator is in 3D.

%In this case the additional constraint of allowing only variations on the 2D plane, leads to define the frame change using only the in-plane magnetic 

As an illustration of how the Lorentz indicator looks in a well known 2D reconnection case,
we consider here a case often used in the literature: a current sheet where the tearing instability \citep{coppi1965current} develops in many sites that coalesce into progressively larger islands \citep{finn1977coalescence}. Specifically we consider a  case of the type reported in \citep{eriksson2014signatures,eriksson2015multiple}: the initial state is a force free equilibrium:
\begin{equation} \label{eq:force free}
\begin{split}
B_x(y) &= B_0 \tanh(y/\delta)\\
B_z(y) &= \Bigl(B_0^2 + B_g^2 - B_x^2 \Bigr)^{1/2}
\end{split}
\end{equation}
with a very narrow current sheet $\delta/d_i=0.1$ (to promote quick short scale tearing) and uniform plasma pressure and density. The plasma properties represent the the solar wind Custer observation reported in where more details can be found \citep{eriksson2015multiple}:  guide field  $B_g/B_0 = 3.9$ and 
temperature $T_e = T_i = 10$ eV.  We use  $m_i/m_e = 256$ and  a domain  $200 \times  30 d_i$. The problem is simulated using the code iPic3D~\citep{ipic3d}\footnote{Available at \url{www.ipic3d.org}} using $2560 \times 384$ cells and 770 billion particles.

As it turns out, the speed of the Lorentz null-B frame is vastly larger than the speed of light in most of the domain: several orders of magnitude. Visualization of a field having values of many orders of magnitude except for a few near singular locations where it drops near zero is visually arduous. We prefer then to renormalize this indicator as:
\begin{equation}
    \mathscr{L} = \log_{10}\left(\frac{v_L}{c}\right)
\end{equation}
The advantage of using the indicator $\mathscr{L} $ is that it varies more smoothly, being zero at the speed of light, positive for superluminal speeds and negative for subluminal speeds.

\begin{figure}[h]
    \centering
    \includegraphics[width=\columnwidth]{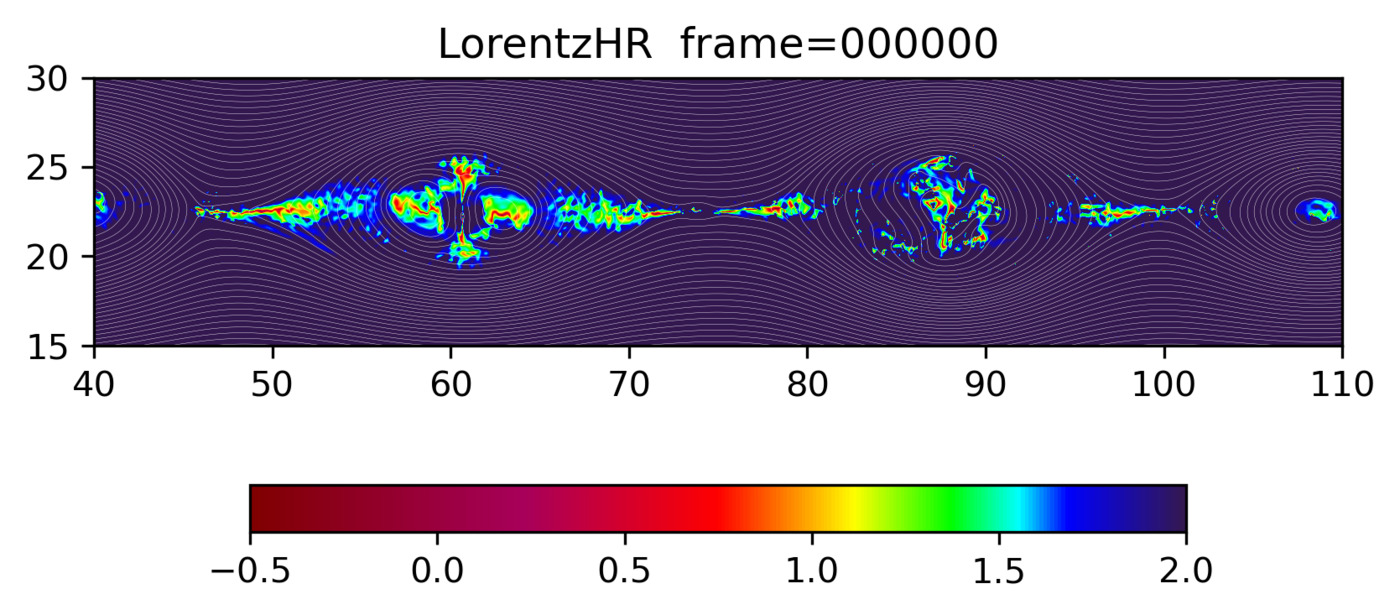}
    \caption{False color representation of the Lorentz indicator $\mathscr{L}$ in a sub-domain $x=[40,110]$ of a simulation of an initial force free current sheet in the solar wind \citep{eriksson2015multiple}.  Superimposed we show the isocontours of the vector potential $A_z$. Note the asymmetric colorscale chose to make the area around small vlaues of $v_L$ more visible. The pixels with $v_L<1$ (amaranth to garnet)  is hardly visible if the surrounding region is not highlighted too.}
    \label{fig:2dforcefreee}
\end{figure}

Figure~\ref{fig:2dforcefreee} shows the Lorentz indicator in a sub-domain (blown up to see more clearly the features). We show two full islands and half of two neighboring ones and three x points in between. As it can be seen the Lorentz indicator identifies all of these feature as well as providing considerable information about the local structure. Strictly speaking the reconnection site proper is defined where $v_L=0$. This statistically happens in between grid points but the grid point closest to it will have the lowest $v_L$. Note that the information of the Lorentz indicator is not just to quantitatively count the reconnection sites but it also provides a qualitative information on how active the reconnection is and how widely it affects its vicinity: more active reconnection regions appear as brighter, wider patches in the indicator.

Figure \ref{fig:2dforcefreee} also show the usual vector potential. In 2D the established procedure is to use the vector potential to find x and o points. It is informative to check what the Lorentz indicator does in the proximity of every topological feature identified by the vector potential. As can be seen, the o-point at the island centers and the reconnection x-points  between magnetic islands are all captured.

As noted above, the proposed Lorentz indicator is no replacement for null point finders. Its aim is different: it wants to find locations where an electric field carries magnetic field lines, breaks them and  pulls them apart. This in 2D can only happen in a x-point (or in a o-point where magnetic field lines can be piled up). In 3D other possibilities arise and in the next section we will see what happens in a 3D simulation.

\section{Detection of reconnection in simulated 3D turbulent reconnection outflows}
We test the reconnection indicator for the Lorentz  null magnetic field frame transformation in a  turbulent outflow of a 3D reconnection region. Previous work has identified that even in initially laminar conditions, the process of reconnection in 3D leads to outflows rich in free energy due to shears, superalfvenic electron beams, thin currents. These sources of energy cause secondary instabilities: drift \citep{divin2015evolution}, Kelvin-Helmholtz\citep{fermo2012secondary}, interchange \citep{nakamura2002interchange}, Buneman \citep{Goldman2008}, (drift)-tearing \citep{daughton2011role}. The combined effect of these instabilities is to make the flow turbulent. The turbulent properties of the outflow have been investigated in previous 3D simulations \citep{pucci2017properties} and shown to agree remarkably with in situ measurements of the turbulent spectrum of electric and magnetic field observed by the Cluster mission \citep{eastwood2009observations}. 

We use here a simulation very similar to that reported in the aforementioned reference by \citet{pucci2017properties}. Previous work has pointed out that the turbulent outflow is the possible host for the formation of secondary reconnection sites \citep{lapenta2015secondary}. Often reconnection affects only electrons but leaving ions unaffected \citep{phan2018electron,stawarz2019properties}. 
Analyses based on kinetic indicators of reconnection, such as slippage, non-ideal terms of the generalized Ohm's law and agyrotropy support that conclusion \citep{lapenta2015secondary}. So far direct inspection of the suspect sites was the main way to investigate whether the points flagged by these indicators were likely to be real reconnection sites or not. This worked in presence of null points, but with difficulty and relying on the painstaking selection of multiple field lines to follow. However, secondary reconnection sites in reconnection outflow are likely to be often unrelated to  nulls and resemble more the twisting of field lines wrapped in flux ropes, as described in \citet{ergun2016magnetospheric}. The visual inspection of field lines is of limited use because at any given time the field lines all form bundles whose internal rotation is linked to reconnection but makes it impossible to observe a field line before and after breaking up and reconnecting. 

The Lorentz subluminal frame change indicator changes this situation. We can compute the frame velocity leading to the elimination of the magnetic field in the plane $(\xi_1, \xi_2)$ and flag the regions where the speed drops below the speed of light. 

To study the application of the Lorentz frame indicator we consider a 3D fully kinetic simulation of reconnection outflows conducted again with the iPic3D PIC code~\citep{ipic3d}. We consider the same type of simulation recently reported in \citep{lapenta2015secondary}: it uses a standard Harris equilibrium \citep{Harris1962}:
\begin{gather}
    \bfB(y)=B_0 \tanh(y/L) \bhx + B_g \bhz \\
    n(y)=n_0\sech^2(y/L)+n_b
\end{gather}
defined by the thickness $L/d_i=0.5$ and with the parameters $m_i/m_e=256$, $v_{the}/c=0.045$, $T_i/T_e=5$. A guide field $B_g/B_0=0.1$ and a background plasma of $n_b/n_0=0.1$ is imposed, where $B_0$ is the asymptotic in plane  field and $n_0$ is the peak Harris density.  To convert code units to physical units we chose as reference: $B_0=20\si{nT}$ and $n_0=\num{1e6}\si{m^{-3}}$.

The coordinates are chosen with  the initial  Harris magnetic field  along $x$ with size  $L_x=40d_i$, the initial gradients  along $y$ with $L_y=15d_i$. The third dimension, where the initial current and guide field are directed,  is initially uniform with $L_z=10d_i$. Open boundaries are imposed in $x$ and $y$ and periodicity is assumed along $z$. 
 
The conditions are identical to what is reported in \citep{pucci2017properties} but the grid resolution is  higher with 1200x450x300  cells, resolving better the electron skin depth, in the reconnecting background plasma $\Delta  x \approx 0.2 d_{e,b}$ (the initial Harris plasma is quickly swept away by reconnection). The time step  resolves also well the electron cyclotron frequency, even in the strongest field, $\omega_{ce,Bmax}\Delta t \approx 0.3$ (and even better in more average fields).  Each plasma species is described by two populations of $5^3$ computational particles per cell, one represents the Harris density ($n_0\sech^2(y/L))$ and the other the background $n_b$. A total of 81 billion particles and 162 million cells is used in a supercomputer parallel topology of 50x30x20 (i.e. 30,000) processors. 

Reconnection is initialized with a uniform perturbation along $z$, very localized in the center of the domain~\citep{lapenta2015secondary}. The evolution of the system is described in details in our previous works~\citep{lapenta2015secondary,lapenta2016energy,pucci2017properties,lapenta2020local}: the initial perturbation quickly produces a  primary reconnection region that initiates the evolution leading to a strong outflow that interacts with the surrounding plasma forming pileup regions. The outflow develops turbulence \citep{vapirev2013formation,divin2015evolution,pucci2017properties,lapenta2020local} and secondary reconnection sites \citep{lapenta2015secondary, liu2018detection}. We report here the conditions at cycle 18000, when we stop the simulation as the pile up region is reaching the boundary and about to exit. 

\begin{figure}
    \centering
    \includegraphics[width=\columnwidth]{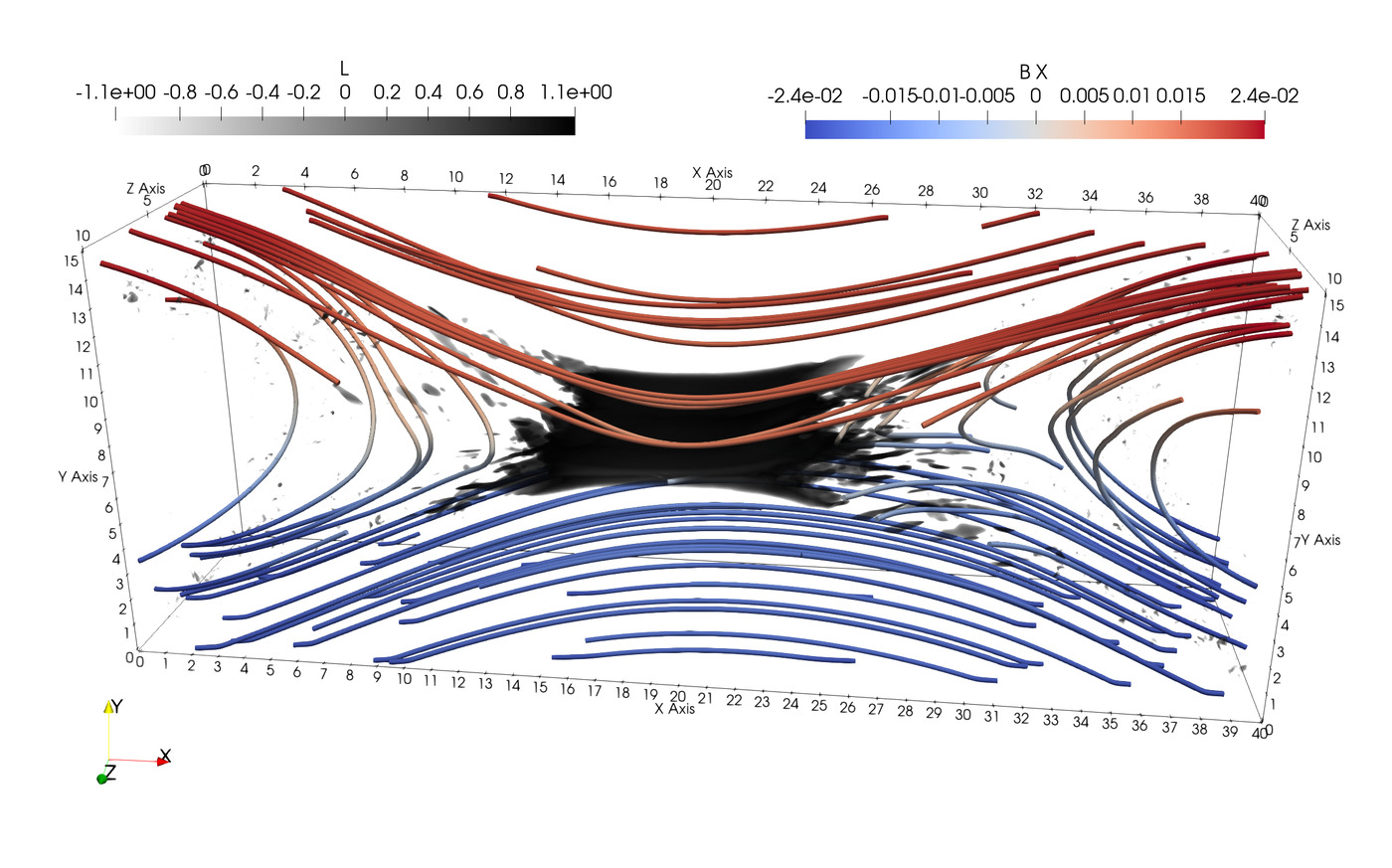}
    \caption{Volume rendering of the Lorentz frame indicator $\mathscr{L}$. A negative value implies subluminal speed for the Lorentz transformation that eliminates the magnetic field in the reconnection plane. A myriad locations  in the outflow as well as the main central reconnection regions are flagged as locations where the Lorentz transformation can eliminate the magnetic field at subluminal speed. }
    \label{fig:volume_lorentz}
\end{figure}

The resulting data-set gives us information about all fields and moments on the points of the simulation grid. In each, we can compute the Lorentz frame indicator $\mathscr{L}$. A volume rendering is shown in Fig.~\ref{fig:volume_lorentz}.
A very large number of tiny locations are identified. Visual inspection is challenging and a hawkish eye is needed to spot the smallest. The color scale used is the best among the many possible to highlight the regions but the task is still not for every eye. Some of the Lorentz subluminal regions shown in Fig.~\ref{fig:volume_lorentz} are indeed just one pixel. 

In fact, in principle, the location of a singular point is a point only, infinitesimally small. 
Obviously the likelihood of one such point sitting exactly on a grid location is vanishingly small, but luckily, as shown in Sect \ref{sect:2dnulls}, the whole neighborhood of a magnetic point with singular topology  has subluminal $v_L$ and we can still identify them by considering all grid points where the speed is subluminal: $\mathscr{L}<0$. 

For this reason, we resort to automatic sorting. We task a computer algorithm with identifying every point on the grid where the Lorentz transformation speed that eliminates the magnetic field becomes subluminal. We consider the regions distinct only if the points are at least one electron skin depth apart and in each region we take the point with the minimal speed. Note that this method takes only into account the grid points. A possible improvement for future work is to  interpolate the fields within each cell and look for the minimum Lorentz speed anywhere within the cell. This is similar to the approach needed to find all null points in a grid-based field: the nulls are within the cells not at the grid nodes~\citep{haynes2007trilinear,fu2015find,Olshevsky:2016ApJ}. For the Lorentz frame indicator the need is less acute because the indicator remains subluminal in an electron skin depth scale region making the identification easier. 
We limit ourselves here only to the grid points because the results do not seem to indicate the need for a more refined sub-grid interpolation. However, this conclusion is valid only in the present example where the electron scale is well resolved. In simulations that do not resolve the electron skin depth well, the grid point-based algorithm can  miss  regions of subluminal indicator that never comprise a grid point and are fully enclosed within a single cell. In that case it becomes necessary to rely on subgrid interpolations to locate the true minimum of $v_L$ within each cell.

In the search, we exclude the central region $x/d_i\in[18,22]$ shown in Fig.~\ref{fig:volume_lorentz} as a large area of subluminal speed. This is the central reconnection region caused by our initial perturbation: we know it is there because we put it there, we look only for the secondary reconnection regions.

The analysis finds in this case 54 secondary points, listed in Table \ref{tab:points}.  Of these, the vast majority are x-like saddle points (39), 6 are attracting foci, 5 repelling foci, 3 are repelling and 1 is an attracting point. There is no perfect center: a center is the limit of attracting and repelling foci when the real part is exactly zero, or in other words when the divergence of the in plane magnetic field component is exactly zero. In 3D, the field will always have a minimum of variation in the third dimension and perfect centers can't be expected.  However, site number 5 is a repelling focus with relatively small real part, resulting in a nearly visual center (see Fig.~\ref{fig:recon_sites2} below).
\begin{figure}
    \centering
    %a)\\
    %\includegraphics[width=\columnwidth]{figure1000}
    %b) \hskip 8cm c)\\
    %\includegraphics[width=\columnwidth]{figure1001}
    \includegraphics[width=.9\columnwidth]{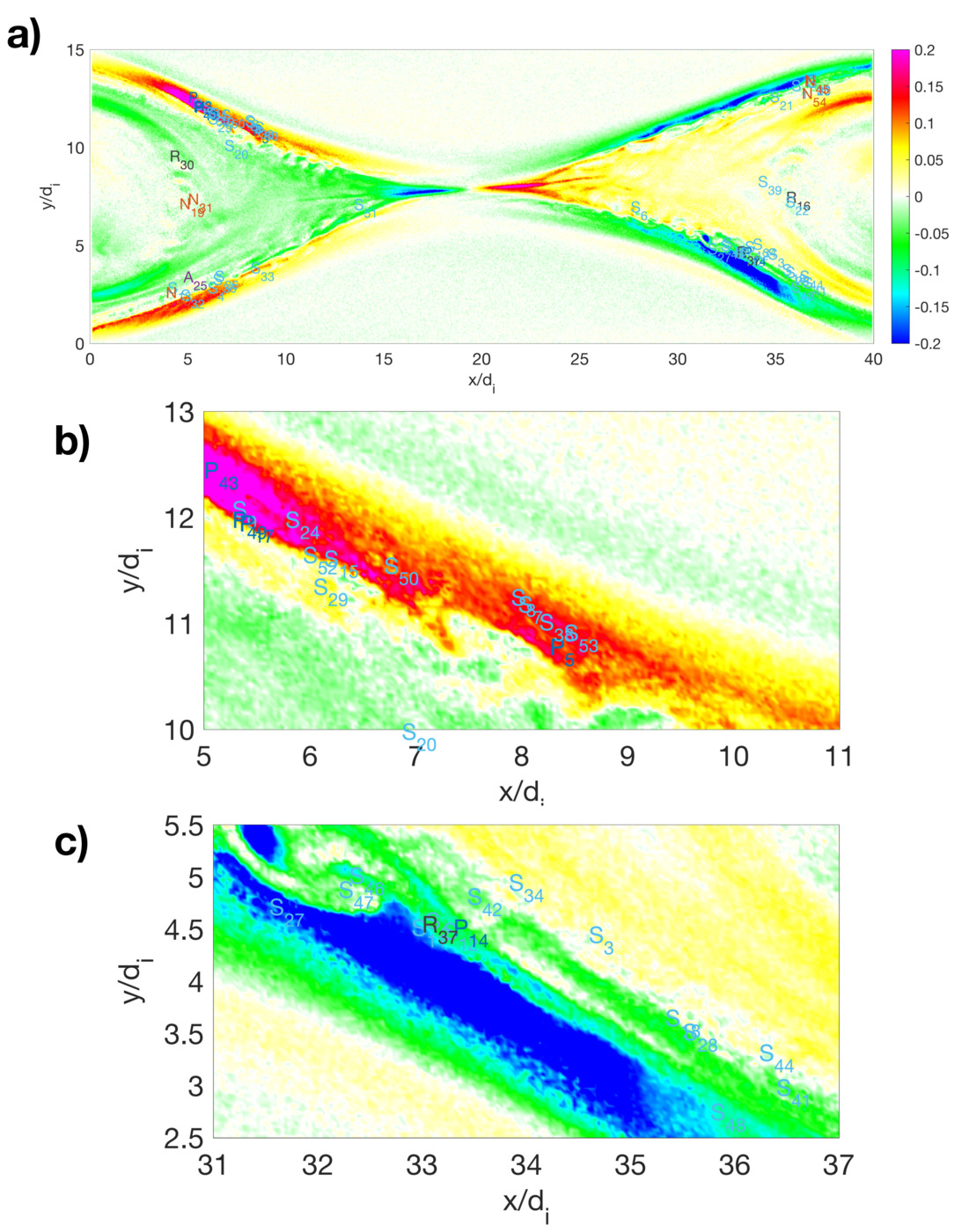}
    \caption{False color rendering of the electron speed $V_{ex}(x,y, z=L_z/2)$ in the mid-plane along $z$. The location and type of the points where the Lorentz frame indicator becomes subluminal is shown, corresponding to the points of singular topology listed in Table \ref{tab:points}. Panel a gives a complete overview and panel b and c blow up the two busiest separatrix regions:top-left and bottom-right. The point detected are numbers as in Table~\ref{tab:points} and marked with their type: center (C), saddle (S), repelling (R), attracting (A), repelling focus (P) and attracting focus (N).Note that the points are not on the plane show and each have a  different $z$ but are projected on the plane shown.}
    \label{fig:sites}
\end{figure}
\begin{table}
\hskip -2cm
\begin{minipage}{.5\linewidth}
\footnotesize
\begin{tabular}{|c|c|c|c|c|c|c|}
\hline
Site & Type & $ x/d_i$ & $ y/d_i$  & $ z/d_i$ & $B_{\xi_3}/B_0$  & $\displaystyle \frac{B_{\xi_3}}{||J_B||_{2}d_i}$  \\ \hline
1 & S &    35.80 &    13.00 &     6.87 &     0.80 &     7.19 \\ 
2 & S &    33.03 &     4.53 &     7.67 &     0.66 &     3.22 \\ 
3 & S &    34.60 &     4.40 &     3.40 &     0.69 &     4.40 \\ 
4 & S &     6.00 &     2.67 &     5.43 &     0.81 &     2.23 \\ 
5 & P &     8.27 &    10.73 &     5.97 &     0.64 &     1.91 \\ 
6 & S &    27.60 &     6.80 &     5.23 &     0.52 &     2.32 \\ 
7 & S &     7.97 &    11.13 &     5.87 &     0.67 &     2.20 \\ 
8 & S &     7.90 &    11.20 &     6.00 &     0.67 &     2.25 \\ 
9 & S &     5.27 &    12.03 &     2.33 &     0.76 &     2.29 \\ 
10 & N &    36.53 &    13.20 &     3.00 &     0.84 &     3.82 \\ 
11 & S &    32.90 &     4.47 &     7.60 &     0.67 &     3.14 \\ 
12 & S &    33.20 &     4.43 &     7.73 &     0.66 &     2.93 \\ 
13 & S &     3.97 &     2.67 &     6.40 &     0.80 &     6.05 \\ 
14 & P &    33.30 &     4.47 &     8.00 &     0.67 &     3.17 \\ 
15 & S &     6.13 &    11.57 &     2.63 &     0.72 &     1.99 \\ 
16 & R &    35.53 &     7.30 &     4.90 &     0.48 &     4.29 \\ 
17 & P &     5.33 &    11.90 &     2.20 &     0.75 &     2.29 \\ 
18 & S &    35.33 &     3.60 &     1.87 &     0.77 &     3.25 \\ 
19 & N &     4.57 &     6.97 &     5.50 &     0.48 &     4.00 \\ 
20 & S &     6.87 &     9.93 &     8.27 &     0.50 &     2.73 \\ 
21 & S &    34.70 &    12.43 &     4.57 &     0.76 &     3.63 \\ 
22 & S &    35.50 &     7.07 &     1.47 &     0.48 &     3.79 \\ 
23 & S &    36.60 &    13.17 &     3.10 &     0.84 &     3.41 \\ 
24 & S &     5.77 &    11.93 &     2.40 &     0.74 &     2.13 \\ 
25 & A &     4.77 &     3.23 &     1.03 &     0.72 &     4.80 \\ 
26 & S &     6.37 &     3.27 &     2.00 &     0.75 &     3.10 \\ 
27 & S &    31.53 &     4.67 &     0.70 &     0.66 &     2.10 \\  \hline
\end{tabular}
  \end{minipage}%
    \begin{minipage}{.5\linewidth}
    \footnotesize
\begin{tabular}{|c|c|c|c|c|c|c|c|}
\hline
Site & Type & $ x/d_i$ & $ y/d_i$  & $ z/d_i$ & $B_{\xi_3}/B_0$ & $\displaystyle \frac{B_{\xi_3}}{||J_B||_{2}d_i}$\\ \hline
28 & S &    35.50 &     3.47 &     7.70 &     0.78 &     2.79 \\ 
29 & S &     6.03 &    11.30 &     5.50 &     0.67 &     2.57 \\ 
30 & R &     4.07 &     9.40 &     2.93 &     0.59 &     3.24 \\ 
31 & N &     5.00 &     7.20 &     8.13 &     0.41 &     1.81 \\ 
32 & S &     4.63 &     2.30 &     2.73 &     0.84 &     2.07 \\ 
33 & S &     8.20 &     3.70 &     5.80 &     0.74 &     1.50 \\ 
34 & S &    33.83 &     4.90 &     5.77 &     0.62 &     2.20 \\ 
35 & S &    36.60 &    13.20 &     3.03 &     0.85 &     3.37 \\ 
36 & S &     6.27 &     3.10 &     1.50 &     0.76 &     2.26 \\ 
37 & R &    33.00 &     4.50 &     8.43 &     0.65 &     1.87 \\ 
38 & S &     8.17 &    10.97 &     6.93 &     0.66 &     1.61 \\ 
39 & S &    34.10 &     8.10 &     7.63 &     0.34 &     1.64 \\ 
40 & N &     3.87 &     2.43 &     5.23 &     0.81 &     7.38 \\ 
41 & S &    36.40 &     2.93 &     1.87 &     0.84 &     2.39 \\ 
42 & S &    33.43 &     4.77 &     5.60 &     0.63 &     1.59 \\ 
43 & P &     5.00 &    12.40 &     7.10 &     0.77 &     1.80 \\ 
44 & S &    36.23 &     3.27 &     5.33 &     0.80 &     2.43 \\ 
45 & N &    36.47 &    13.27 &     2.87 &     0.84 &     1.44 \\ 
46 & S &    32.30 &     4.97 &     8.03 &     0.61 &     1.42 \\ 
47 & S &    32.20 &     4.83 &     8.00 &     0.62 &     1.37 \\ 
48 & S &    35.77 &     2.70 &     7.93 &     0.82 &     1.44 \\ 
49 & P &     5.27 &    11.93 &     2.20 &     0.76 &     1.14 \\ 
50 & S &     6.70 &    11.50 &     5.07 &     0.71 &     0.94 \\ 
51 & S &    13.47 &     6.93 &     6.97 &     0.28 &     0.82 \\ 
52 & S &     5.93 &    11.60 &     8.93 &     0.71 &     1.10 \\ 
53 & S &     8.40 &    10.87 &     7.27 &     0.65 &     0.54 \\ 
54 & N &    36.33 &    12.60 &     4.17 &     0.77 &     1.08 \\  \hline
\end{tabular}
  \end{minipage}%
    \caption{Singular topological points identified by the Lorentz transformations as locations were a subluminal transformation eliminates the field $B_\perp$. The points are numbered as in the Fig. \ref{fig:sites} their type and location is provided. The type is indicated as center (C), saddle (S), repelling (R), attracting (A), repelling focus (P) and attracting focus (N), for the types illustrated in Fig.~\ref{fig:sites}}
    \label{tab:points}
\end{table}

The positions of the topological points identified by the Lorentz transformations are projected in the $(x,y)$ plane of the simulation in Fig.~\ref{fig:sites}. As noted above, we have excluded the points in the central region $x/d_i\in[18,22]$. As can be seen, the points are concentrated in the separatrix region where the inflow and outflow interact. The points are located both in the flow region going towards the primary reconnection site (yellow-red on the left side and green-blue on the right side) and the outflow from the reconnection site (opposite colors). This is a region of strong turbulence \citep{lapenta2020local} caused by drift modes and Kelvin-Helmholtz instability due to the presence of density and velocity gradients  \citep{Divin2012} and by drift tearing modes due to the strong thin current layers in the region \citep{daughton2011role}. Note that the outflow from reconnection remains collimated only in the central region \citep{goldman2011jet} but turns to a  turbulent flow that fills the domain around the pile up region in the outflow. The presence of a mild but significant guide field in this simulation explains the bend in the outflowing jet \citep{goldman2011jet} and the asymmetry between the upper and lower separatrices \citep{kleva1995fast,gembeta,birn-priest}.

\section{Analysis of the detected  sites.}

\begin{figure}
    \centering
    %\begin{tabular}{ccc}
    %\includegraphics[width=.3\columnwidth]{recon_site000001}& \includegraphics[width=.3\columnwidth]{recon_site000014} &
    %\includegraphics[width=.3\columnwidth]{recon_site000019}
    \includegraphics[width=.9\columnwidth]{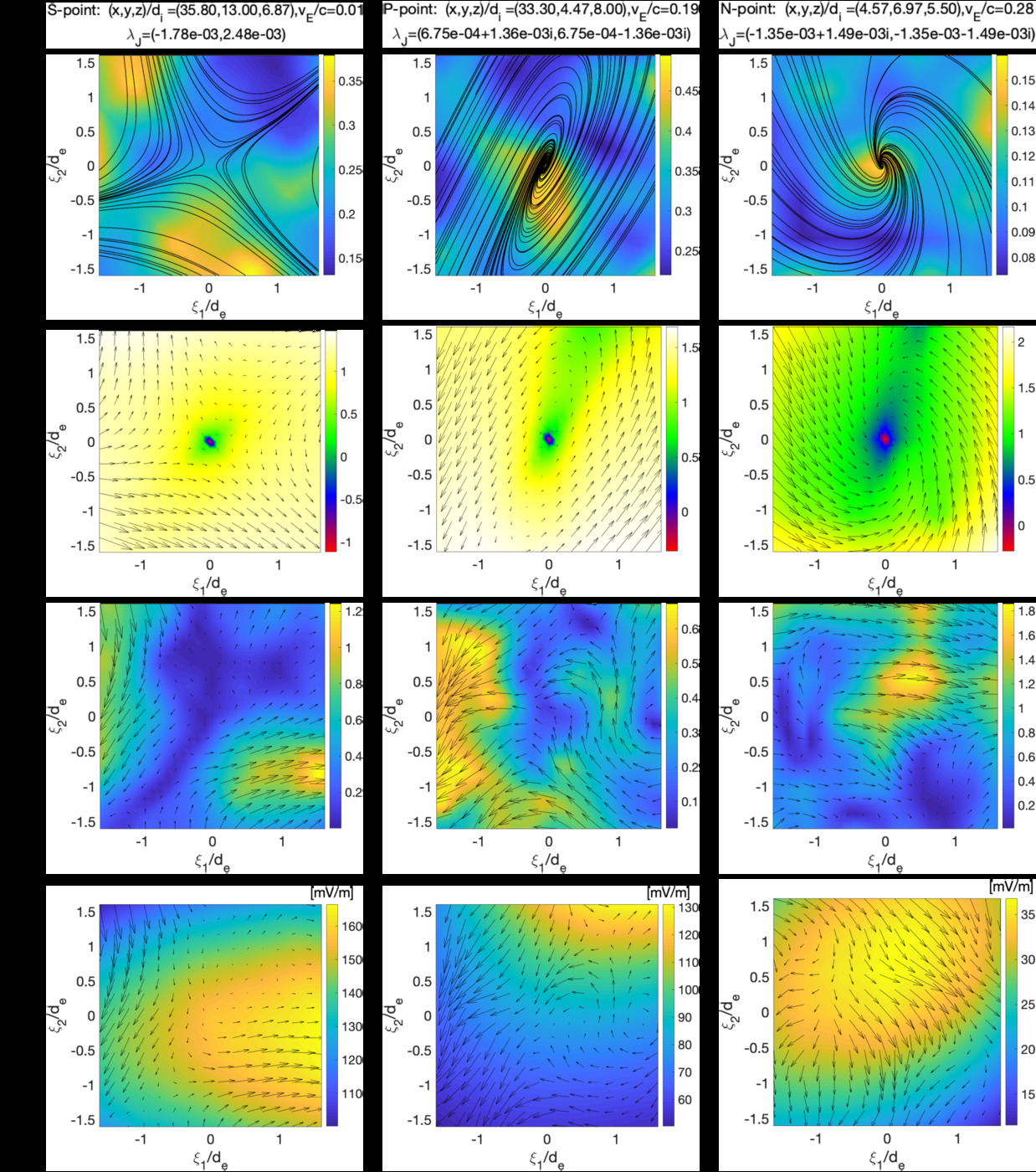}
    %\end{tabular}
    \caption{Conditions around three types of detected points: a saddle (S, site 1), and repelling (P, site 14) and an attractive (N, site 19 focus). For each column corresponding to a different point, four panels are shown from top to bottom. Top row: in plane magnetic field lines superimposed over the electron agyrotropy of the electron pressure tensor, $\mathcal A$. Second row:  False color representation of the Lorentz frame indicator $\mathscr{L}$  with superimposed arrow plot of the in plane magnetic field, $\bfB_{(\xi_1,\xi_2)}$. Third row, false color and arrow plot of the electron flow  velocity $\bfv_{e(\xi_1,\xi_2)}$ (the color indicates the modulus of the speed normalized by the speed of light). Fourth row, false color representation of the reconnection electric field $E_{\xi_3}$, with superimposed arrow plot of the in plane relative electron-ion flow velocity, $\bfv_{e(\xi_1,\xi_2)} -\bfv_{i(\xi_1,\xi_2)}$. The data reported here is provided for all 54 detected sites in the form of a movie where each frame contains the same four panels shown here: \href{run:supplemental/combo.mp4}{Video}}
    \label{fig:recon_sites1}
\end{figure}

\begin{figure}
    \centering
    %\begin{tabular}{ccc}
    %\includegraphics[width=.3\columnwidth]{recon_site000016}& \includegraphics[width=.3\columnwidth]{recon_site000025} & \includegraphics[width=.3\columnwidth]{recon_site000005}
    \includegraphics[width=.9\columnwidth]{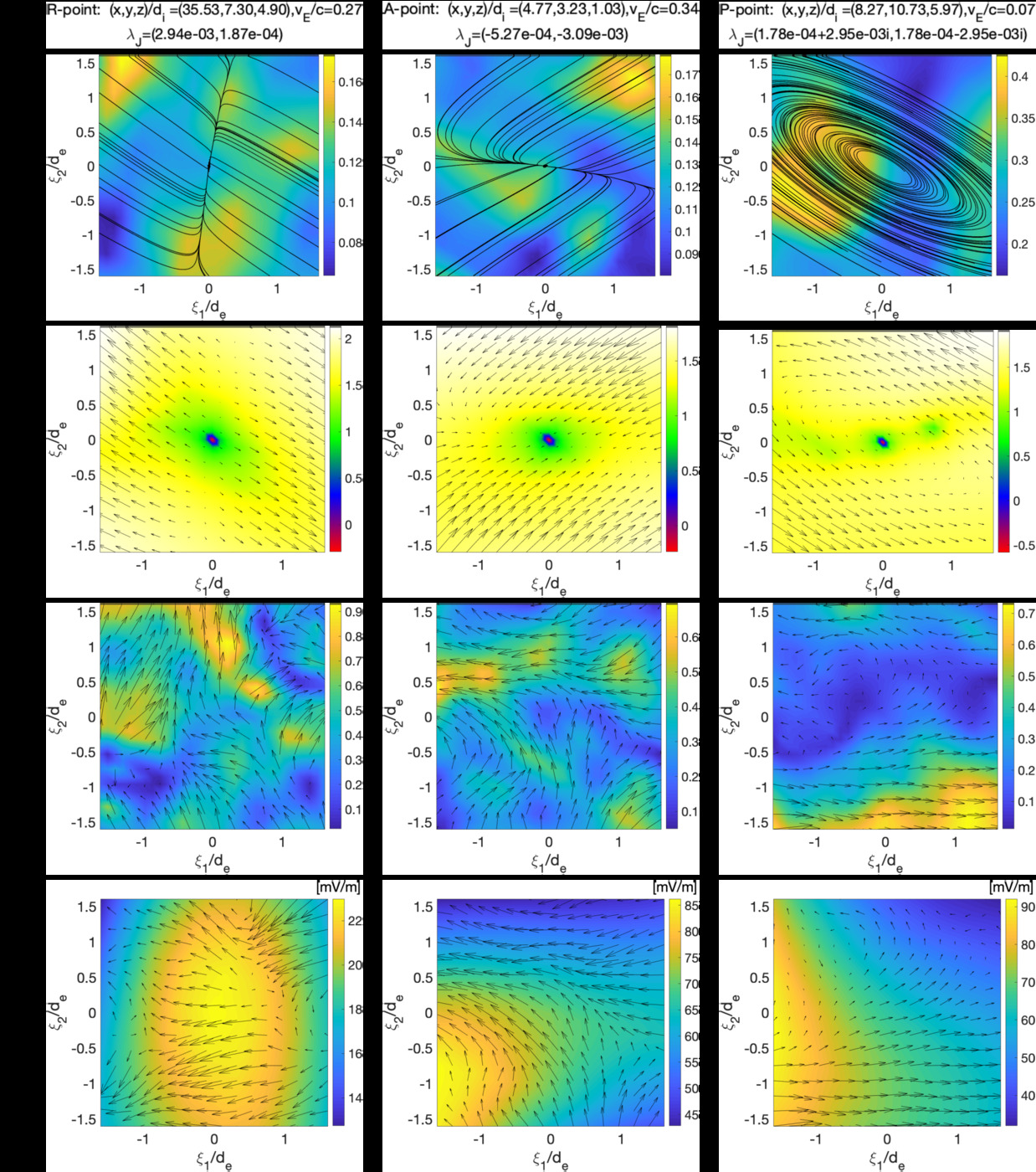}
%    \end{tabular}
    \caption{Conditions around three types of detected points: a repelling (R, site 16), an attractive focus (A, site 25) point and a repelling focus (P, site 5) that is in fact nearly a center as the spiraling dominates over the outward going of the field lines. The four rows have the same type of information as in Fig.~\ref{fig:recon_sites1}}
    \label{fig:recon_sites2}
\end{figure}

The method described above identified 54 potential points of interest where reconnection might be involved. A detailed analysis  of all 54 of them is provided in the supplemental material as a movie where each frame is one detected point. Let us focus on a few of the most characteristic cases. Figures~\ref{fig:recon_sites1}-\ref{fig:recon_sites2} show all six cases illustrated in Fig.~\ref{fig:sites} and listed in Table \ref{tab:points}. 

Some features are common to all sites. 

First, there is always right on spot or in very close proximity  a significant agyrotropy. Agyrotropy signifies a complex structure of the pressure tensor with significant off-diagonal terms and has come to be a  reliable indicator of reconnection \citep{scudder2008illuminating,aunai2013electron,swisdak2016quantifying}.  The off-diagonal terms of the pressure tensor contribute non ideal terms to the generalized Ohm's law and allow the violation of the frozen in condition \citep{biskamp,birn-priest,hesse-guide,gembeta}. A key discovery of the Magnetospheric MultiScale (MMS)  mission was the observational detection a non gyrotropic electron pressure tensor, confirming the theoretic prediction \citep{burch2016electron}. This is a critical need to allow electrons to slip with respect to the field lines, which is central to the definition of reconnection. Note that agyrotropy is typically highest just outside a reconnection site and not quite at it \citep{scudder2008illuminating,burch2016electron}.

%Second, the density is usually non uniform but not dramatically so if one looks carefully at the color scale used. It is still possible to interpret these sites in terms of symmetric-type reconnection or at most weakly asymmetric in a few cases. But clearly we are not seeing a strong local discontinuity in the density as is for example the case the of electron holes or separatrix cavities. 

Second, the Lorentz frame indicator $\mathscr{L}$ is minimum at the selected locations, by construction of the selection method, but the speed of the Lorentz transformation that eliminates the magnetic field  remains subluminal (i.e. the indicator $\mathscr{L}<0$) in a non singular and smooth area around the point measures, allowing for an easier identification.

Third, in the vicinity of the identified sites there is a significant reconnection electric field. By definition we choose our coordinate system to align the $\xi_3$-axis with the reconnection electric field and with the guide magnetic field. The reconnection electric field reported in SI units measures the strength of the reconnection rate.  It is in fact perfectly possible to have a quiescent magnetic singularity, one not accompanied by a significant electron flow around it, this is a potential reconnection site that is at the moment not reconnecting \citep{olshevsky2015energy}. As can be seen, however, the reported cases all show significant reconnection electric fields, in fact a field of \num{100}\si{mV/m} is exceptionally high when compared with in situ data reported by MMS \citep{ergun2016magnetospheric}.

Fourth, as shown in the table \ref{tab:points}, all points have a very substantial guide magnetic field. This is supported by the recent observation done by MMS where  electron-scale secondary reconnection sites are observed to have preferentially a sizable guide field \citep{stawarz2019properties}. This might be here more a consequence of the fact that the domain has no large scale region with small field and reconnection happens where it can. Were there regions of low magnetic field perhaps reconnection would happen also at low guide field. The typical outflow from a primary reconnection site is magnetized for the electrons except for the primary reconnection site itself, so one must expect relatively large guide field in secondary reconnection sites embedded in the magnetic field of the outflow. We mean by large, here, larger than 1/10 the reconnecting field. 

Finally but most importantly, the electron flow around these sites is clearly strong and singular. The reconnection site forms a pattern of flow that disrupts the nearby pattern and determines around it a flow that is clearly aware of the presence of the reconnection site and is affected by it. Detecting the electron flow around a reconnection site is the most common way to identify reconnection. By this standard then the identified sites break the frozen-in condition.  
Not all selected sties show a trivially interpretable flow pattern. This is due to the 3D nature of the flow that includes a out of plane component not shown in the figure. In all sites (see also movie in the supplemental section), the flow is primarily carried by the electrons. We show in the panels on the third row the electron flow speed in the reconnection plane ($\bfv_{e(\xi_1,\xi_2)}$) and in the fourth row the relative electron-ion speed ($\bfv_{e(\xi_1,\xi_2)}$-$\bfv_{i(\xi_1,\xi_2)}$). Obviously in all cases the ions are not moving in a significant manner compared with the electrons, compatible with the secondary reconnection sites being site of electron-scale reconnection \citep{phan2018electron}.

If these are general properties, each site has its own peculiarities best appreciated by looking at all 54 of them in the movie provided in the supplemental material. Some sites are more related to 2D x-points, some more to 2D o-point. In the first case, one can consider the location a reconnection site proper while the latter can be considered the center of a flux rope. 
One case sticks out among the 54: P site 25, shown in Fig.~\ref{fig:recon_sites2} where the real part of the eigenvalues while not hard zero is relatively small allowing us to consider this focus point almost as a center.

\begin{figure}
    \centering
    \includegraphics[width=\columnwidth]{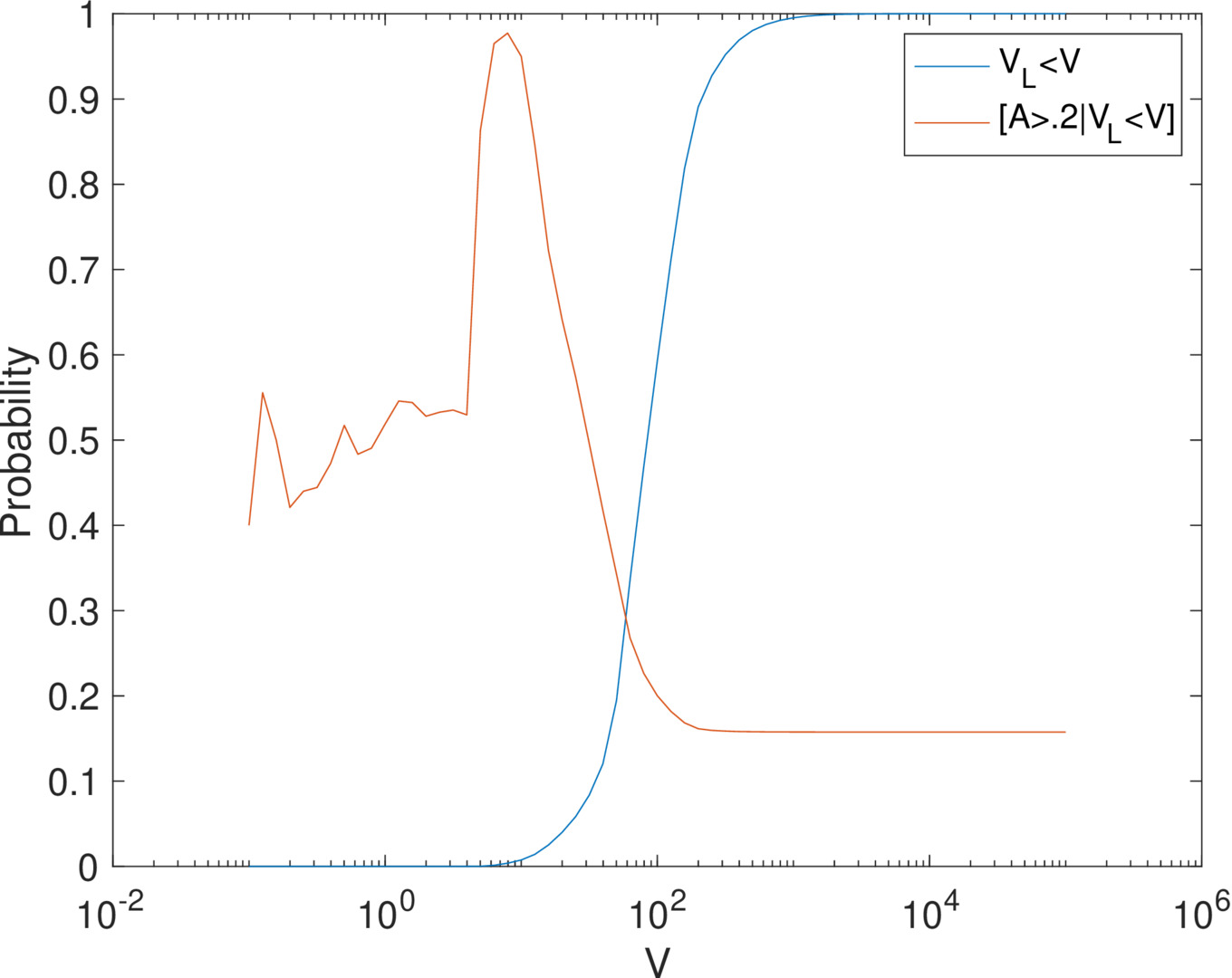}
    \caption{Cumulative probability distribution of the velocity $v_L$ of the local Lorentz frame transformation that eliminates the magnetic field,  indicating the probability of finding a pixel with speed $v_L$  below a certain value $V$ (blue). Conditional probability of finding a pixel with agyrotropy exceeding a threshold value of 0.2 among the points with $v_L<V$ (red).}
    \label{fig:cdf}
\end{figure}

The evidence provided partially in Figs.~\ref{fig:recon_sites1}-\ref{fig:recon_sites2} and completely in the movie in the supplemental material is that there is always a substantial agyrotropy in the vicinity of a detected point, as observed also in  in-situ studies~\citep{chen2019electron}.  Figure \ref{fig:cdf} makes this assessment quantitative. We report the cumulative probability of finding a point with $v_L$ less than a certain value $V$. As can be seen the probability of finding subluminal points is exceedingly small: topological singularities are a rare find. To be precise, in the simulation reported, the probability of finding a pixel with subluminal Lorentz frame speed $v_L$ is \num{2.1296e-06} (approximately the same of a royal flush in the game of poker). This is then also the probability of finding a reconnection site. Other  methods that do not use the Lorentz frame indicator have to face these adverse odds. A spacecraft flying in situ will also have a similar hard time finding precisely such points. The proposed Lorentz frame indicator, instead, finds them naturally by construction and it can detect also the neighborhood where the speed while not subliminal drops by several orders of magnitude compared with the rest of the domain, signaling a nearby reconnection site.

Figure \ref{fig:cdf} additionally shows the conditional probability that  the agyrotropy in the points with $v_L<V$ is larger than 0.2 (a value to be considered quite substantial as the typical maximum value observed near a reconnection site is 0.4~\citep{scudder2008illuminating}, we are using the half maximum value). As can be seen this conditional probability reaches nearly one for small values of V but then drops to only 50\% when the selectivity of $v_L$ becomes small. The meaning is simple: there is in the vicinity of singular point a significantly high agyrotropy but only in 50\% of the cases the high agyrotropy is right on the point. This is not unexpected because the agyrotropy is large in the vicinity of a reconnection site but not exactly at it \citep{scudder2008illuminating}.

\section{Conclusions}
Finding reconnection sites in 3D is a notoriously difficult problem. Part of the difficulty is that it is difficult to even come to a universal definition of what 3D reconnection is. 

We take here a pragmatic definition that is suitable for the type of reconnection one is interested in finding in a modern large scale fully kinetic simulation. We define reconnection as the process where a reconnection electric field produces a  drift that carries two components of the magnetic field towards a point where they are eliminated and new field lines are formed. 

Once we take this pragmatic definition, we observe that in the well understood 2D case of x and o point, this condition is accompanied by a peculiar property of the Lorentz transformations of the electromagnetic fields under the condition of eliminating the magnetic field in the plane normal to the electric field in the laboratory frame: at the o and x points the speed of the Lorentz transformation that eliminates the magnetic field is zero and in the proximity of it it drops to subluminal speeds. Everywhere else in the domain the speed of this transformation exceeds the speed of light, signifying that there is no transformation that eliminates the normal component of the magnetic field. This observation makes the detection of x and o point very simple: in a simulation they can be found by looking for the points where the speed of this specific type of Lorentz transformation becomes subluminal (i.e. the transformation is possible). 

In 2D this condition can be studied analytically (see Sect.  \ref{sect:2dnulls}) in the vicinity of o and x points and the extension to 3D is straightforward. 

Based on this, we propose an indicator of magnetic singular points where reconnection defined as above takes place. The indicator is given in eq. (\ref{vE}) and can be computed locally, without needing any field aligned integration or any local differentiation.   This indicator can be computed cell by cell and in situ observations point by point.  When a point of  indicator $\mathscr{L}<0$ is identified, the local type of magnetic singularity can be identified studying the Jacobian of the magnetic field. This operation requires non local information to compute the derivatives: This poses  no trouble in a simulation but for in situ observations it requires the use of multi-spacecraft missions where the use of the curlometer technique allow the computation of gradients \citep{dunlop2008curlometer}. The approach is currently used to find magnetic nulls \citep{fu2015find} and it could be extended for the purpose proposed here. However, unlike finding nulls~\citep{haynes2007trilinear,fu2015find} or computing the Poincare-index \citep{xiao2006situ} that require only the magnetic field, the Lorentz indicator is based also on the electric field. The electric field tends to be more noisy. This did not prove to be a issue in the simulation cases studied here but it might be a consideration for application to in situ measurements. 

We tested the approach for  highly resolved 2D and 3D fully kinetic  simulations of  reconnection. We find numerous detected sites and we studied the properties of each potential reconnection site. We find that in virtually 100\% of them there are other concurring indicators of reconnection, in particular a large agyrotropy.

The proposed approach is not meant to replace the well established method of location of magnetic nulls \citep{haynes2007trilinear,fu2015find}. Nulls are just one way in which reconnection can happen in 3D. Our approach is different and aims at finding locations where an electric field is carrying 2 components of B to an annihilation point. Active reconnecting nulls have this property too and indeed the proposed indicator finds the nulls as well because the nulls are a limit case of the approach proposed for the case where not just 2 but all 3 components of the magnetic field vanish at one point. However, the proposed indicator finds also other types of reconnection, unrelated to nulls.

%The Lorentz frame indicator is valid in 3D because only in 3D any frame velocity direction is allowed. In 2D, the Lorentz frame indicator is valid only if the frame velocity is in the same plane of the reconnection process. As noted in Appendix A, this is valid only in a very close vicinity of a reconnection site where there is no in-plane electric field. The in-plane electric field is absent in the extreme vicinity of a reconnection site, within the electron diffusion region but is present just outside it~\citep{chen2011inversion}. In that case the frame speed moves out of the plane considered, invalidating the assumptions. But this limitation is not critical as the Lorentz frame indicator is intended for 3D systems, the 2D case is already well covered by the method based on the out of plane vector potential~\citep{servidio2009magnetic}. 

The pragmatic definition of reconnection used here might not satisfy all proposed definitions of reconnection: It is meant instead to find regions where the processes we have become accustomed to associate with fast kinetic reconnection take place. With this goal in mind we show for one 3D simulation the ability of the method to identify numerous locations where such processes happen: strong electron drift in near complete absence of ion motion (electron scale reconnection), electron temperature agyrotropy and the presence of a reconnection electric field in the direction normal to the plane of the two reconnecting magnetic field components. 

While our aim is operational, the physics interpretation of the Lorentz frame indicator is tantalizing.  Magnetic field lines do not exist and are a figment of our imagination but if we want to give them an existence, they would be moving with the frozen in plasma at the local $\bfE \times \bfB$. We find that in proximity of  reconnection, a frame exists where an observer moving in the same direction of the magnetic field lines  could eliminate the magnetic field without exceeding the speed of light while remaining in the same plane normal to the laboratory electric field. The existence of such a frame appears to be intrinsically linked with the presence of reconnection.

\section*{Acknowledgments}
The author is grateful to John Finn for stimulating discussions on magnetic topology and 3D reconnection and with Fabio Bacchini for the illuminating discussion on Lorentz frame transformations.
This project has received funding from the KULeuven Bijzonder Onderzoeksfonds (BOF) under the C1 project \emph{TRACESpace}, from the European Union's Horizon 2020 research and innovation programme under grant agreement No. 776262 (AIDA) and from the NASA grant 80NSSC19K0841. Computing has been provided
by NASA at the NAS and NCCS high performance computing facilities,
by the Flemish Supercomputing Center (VSC) and by the
PRACE Tier-0 program. This
research used resources of the National Energy Research
Scientific Computing Center, which is supported by the Office
of Science of the US Department of Energy under Contract No.
DE-AC02-05CH11231.

%\bibliography{lapenta}% Produces the bibliography via BibTeX.

\end{document}